\documentclass[dvips]{article}

\usepackage{amsmath,amssymb}
\usepackage{graphicx}
\usepackage[left=2.5cm,top=2cm,right=2.5cm,bottom=1.5cm,nohead,foot=1cm]{geometry}

\begin{document}

\title{A cryogenic beam of refractory, chemically reactive molecules with expansion cooling}
\author{Nicholas R. Hutzler,\footnote{email: hutzler@physics.harvard.edu}\:\footnote{Harvard University Physics Department, Cambridge, MA} Maxwell Parsons,$^\dag$ Yulia V. Gurevich,$^\dag$ Paul W. Hess,$^\dag$ Elizabeth Petrik,$^\dag$\\ Ben Spaun,$^\dag$ Amar C. Vutha,\footnote{Yale University Physics Department, New Haven, CT}\; David DeMille,$^\ddagger$ Gerald Gabrielse,$^\dag$ and John M. Doyle$^\dag$}
\maketitle

\begin{abstract}
 
Cryogenically cooled buffer gas beam sources of the molecule thorium monoxide (ThO) are optimized and characterized. Both helium and neon buffer gas sources are shown to produce ThO beams with high flux, low divergence, low forward velocity, and cold internal temperature for a variety of stagnation densities and nozzle diameters. The beam operates with a buffer gas stagnation density of  $\sim 10^{15}-10^{16}$ cm$^{-3}$ (Reynolds number $\sim1-100$), resulting in expansion cooling of the internal temperature of the ThO to as low as 2 K.  For the neon (helium) based source, this represents cooling by a factor of about 10 (2) from the initial nozzle temperature of about 20 K (4 K).  These sources deliver $\sim 10^{11}$ ThO molecules in a single quantum state within a 1-3 ms long pulse at 10 Hz repetition rate. Under conditions optimized for a future precision spectroscopy application \cite{Vutha2010}, the neon-based beam has the following characteristics: forward velocity of 170 m s$^{-1}$, internal temperature of 3.4 K,  and brightness of $3\times 10^{11}$ ground state molecules per steradian per pulse.  Compared to typical supersonic sources, the relatively low stagnation density of this source, and the fact that the cooling mechanism relies only on collisions with an inert buffer gas, make it widely applicable to many atomic and molecular species, including those which are chemically reactive, such as ThO. 
\end{abstract}

\section{Introduction}

Atomic and molecular beams are important tools for precision spectroscopy and collision studies \cite{Ramsey1985,Scoles1988}.  The most commonly used beam methods are effusive and supersonic (typically pulsed in the latter case).  Effusive beams of certain species have very large fluxes, especially metal atoms or low-reactivity molecules with high vapor pressure.  Often, however, these beams have large velocity spreads and broad distributions over internal states, reducing the useful signal for many spectroscopy experiments.  Supersonic beams, on the other hand, are cold translationally and internally, but have a large forward velocity.  Typically, for spectroscopy, it is desirable to have a beam that is both cold (to narrow spectral features, and, in the case of molecules, concentrate population in a small number of rotational levels) and slow (to increase the interaction time and decrease time-of-flight broadening) while maintaining high flux. There are a range of techniques to achieve such a beam.  For example, beams of many atomic species can be laser-cooled and slowed using dissipative optical forces \cite{Phillips1985}, through techniques such as frequency chirping \cite{Ertmer1985}, Zeeman slowing \cite{Prodan1985}, and white-light slowing \cite{Zhu1991}.  There is growing interest \cite{Carr2009} in extending such slowing and cooling methods to molecules, especially polar molecules, whose rich internal structure and strong dipolar interactions make them candidates for quantum computing \cite{DeMille2002}, cold chemistry \cite{Krems2005}, precision measurement \cite{Vutha2010,Hudson2002,Flambaum2007,DeMille2008}, and observation of novel quantum phases \cite{Micheli2006}.  Laser cooling of a molecule has recently been demonstrated \cite{Shuman2009,Shuman2010}, and other existing techniques to produce cold and slow beams of polar molecules include using time-varying electric fields to decelerate a supersonic expansion \cite{PhysRevLett.83.1558}, filtering slow molecules from a cold source \cite{Rangwala2003}, and buffer gas cooling \cite{Krems2009} of both pulsed \cite{Maxwell2005} and continuous \cite{Patterson2007} beams.

In this paper we report on a newly developed buffer gas cooled beam source.  We study ThO beams from both helium and neon buffer-gas-based beam sources, and observe cooling of ThO from the free expansion of the buffer gas.  The flux, divergence, temperature, and velocity of the buffer gas beams we studied compare favorably (and for the case of chemically reactive molecules \cite{Ths+02}, very favorably) to supersonic or effusive beams, as discussed further in section \ref{sec:otherbeams}.  Buffer gas beams have been successfully utilized in the past to create cold and slow beams of several species \cite{Krems2009,Prd09}, including (but not limited to) ND$_3$, O$_2$, PbO, SrO, ThO, Na, Rb, and Yb.  The beams presented in the current work operate at a much higher Reynolds number than previously demonstrated buffer gas beams, and we observe new and advantageous features, in particular expansion cooling, not seen in previous work.

Both helium and neon cooled beams were studied.  Helium gas can be used at lower temperatures, and can therefore can be used to lower the initial ThO temperature below that of neon-based sources.  However, neon has much simpler pumping requirements to maintain good vacuum in the beam region, and a neon-based source operates at a higher temperature which allows for larger heat loads for molecule production.  We find that due to expansion cooling in neon, the final, optimized beam temperatures for both helium- and neon-based beams are very similar. For a planned precision spectroscopy experiment with ThO \cite{Vutha2010}, the neon-based source is superior in certain aspects (particularly technical ones) to that of the helium-based source.  Further comparison of helium and neon-based sources can be found in section \ref{sec:hevsne}.

\section{Apparatus}\label{apparatus}

The heart of our cold beam apparatus (see figures 1,2) is similar to that which is described in \cite{Maxwell2005}. It is a cryogenically cooled, cylindrical copper cell with internal dimensions of 13 mm diameter and 75 mm length. A 2 mm inner diameter tube entering on one end of the cylinder flows buffer gas into the cell. An open aperture (or nozzle) on the other end lets the buffer gas spray out as a beam, as shown in figure \ref{fig:buffer_cell}. ThO molecules are injected into the cell via ablation of a ceramic target of ThO$_2$, located approximately 50 mm from the exit aperture. A pulsed Nd:YAG laser \cite{theminilite} is fired at the ThO$_2$ target, creating an initially hot plume of gas-phase ThO molecules (along with other detritus of the ablation process). Hot ThO molecules mix with the buffer gas in the cell, and cool to near the cell temperature, typically between 4 and 20 K.  The buffer gas is flowed continuously through the cell at a rate $f_0 = 1-100$ SCCM (standard cubic centimeters per minute, \cite{sccmdef}). This both maintains a buffer gas stagnation density of $n_0\approx 10^{15}-10^{16}$ cm$^{-3}$ ($\approx 10^{-3}-10^{-2}$ torr) and extracts the molecules out the aperture into a beam.  The result, due to the pulsed introduction of ThO into the cell, is a pulsed beam of ThO molecules (embedded in a continuous flow of buffer gas) over a 1-3 ms period. We have achieved stable operation of the neon based beam with a 200 Hz repetition rate, however the data presented in this paper is at a repetition rate of 10 Hz.  The cell aperture is a square hole of adjustable side length $d_\mathrm{a} = 0-4.5$ mm that can be varied \emph{in situ} and continuously while the beam runs.

Either helium or neon is used as the buffer gas (typically called the carrier gas in most beam literature).  The cell temperature $T_0$ is maintained at $5\pm 1$ K for helium, and $18\pm 1$ K for neon.  The cell is surrounded by a radiation shield at 4 K, which is partially covered in activated charcoal \cite{charcoal} to form a cryopump that keeps the helium background pressure low.  The radiation shields and cell are cooled by a pulse tube refrigerator \cite{thecryomech}. The ThO$_2$ target is constructed from ThO$_2$ powder, pressed and sintered as described in \cite{Balakrishna1988}.  These targets yield $>$30,000 YAG shots on a single focus site before the yield per shot drops to 50\% of the initial value, at which time the focus must be moved to a new spot. The large surface area ($\sim$1 cm$^2$) of the target should allow for $>10^7$ shots before target replacement is necessary. 
\begin{figure}[htbp]
	\centering
		\includegraphics[width=0.80\textwidth]{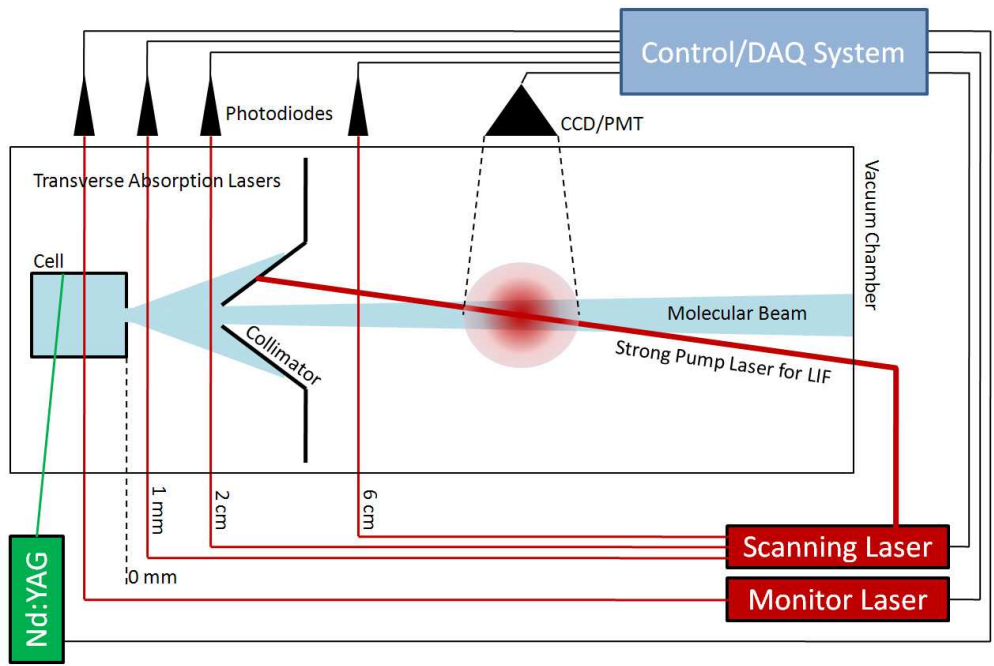}
	\label{fig:apparatus}
	\caption{The cold beam apparatus.  The features inside the vacuum chamber are described in section \ref{apparatus}, and the features outside the vacuum chamber are described in section \ref{results}.  A detailed view of the cell is shown in figure \ref{fig:buffer_cell}.}
\end{figure}

The beam exiting the cell is incident on a conical collimator with a 6 mm diameter orifice located 25 mm from the cell aperture.   There are expected to be few collisions this far from the cell aperture, so, for our work, the features of this collimator are more akin to those of a simple differentially pumped aperture. For helium buffer gas, the collimator is at a temperature of about 8 K. For neon buffer gas the collimator is heated to 30 K to prevent neon ice formation.

\begin{figure}[htbp]
	\centering
		\includegraphics[width=0.50\textwidth]{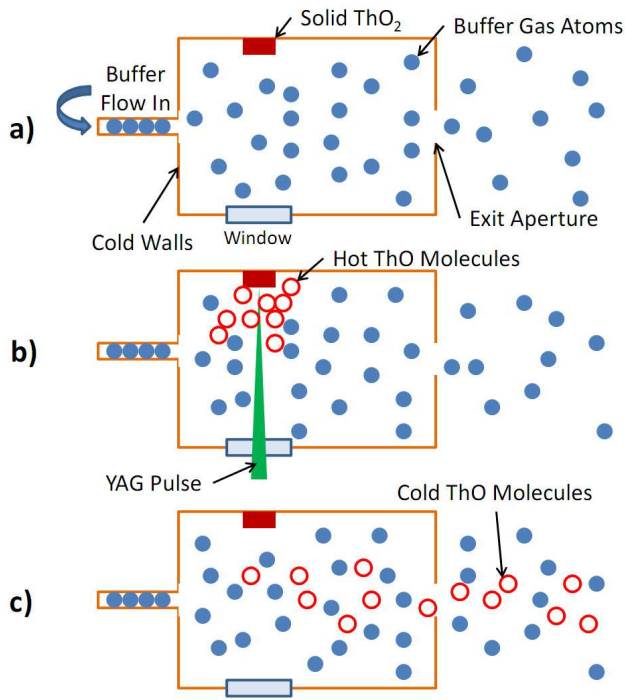}
	\label{fig:buffer_cell}
	\caption{A schematic of the buffer gas cell.  (a) A continuous gas flow maintains a stagnation density of buffer gas atoms (filled blue circles) in thermal equilibrium with a cold cell.  A solid piece of ThO$_2$ is mounted to the cell wall.  (b)  A YAG pulse vaporizes a portion of the target and ejects hot ThO molecules (empty red circles).  (c)  Collisions with the buffer gas cool the molecules, which then flow out of the cell to form a beam.}
\end{figure}

\subsection{Qualitative Features of Buffer Gas Beams}\label{buffertheory}

Here we present a brief overview of the features of a buffer-gas cooled beam, which are discussed in detail in existing literature \cite{beamrefs}. Central to understanding the unique properties of buffer gas cooled beams is the process of hydrodynamic entrainment \cite{Patterson2007}.  A ``hydrodynamic" buffer gas cooled beam is designed so that the characteristic pumpout time $(\tau_\mathrm{pump})$ for the molecules to exit the cell is less than the characteristic diffusion time to the cell walls $(\tau_\mathrm{diff})$ \cite{effusivenote}, both of which are typically 1-10 ms.  These conditions result in many of the molecules being extracted from the cell before they diffuse and stick to the cold cell walls.  The ratio of these timescales is given by
\begin{equation}
\gamma_\mathrm{cell}\equiv\frac{\tau_\mathrm{diff}}{\tau_\mathrm{pump}} \approx \frac{f_0\sigma}{\bar{v}_{0,\mathrm{b}}L_\mathrm{cell}}, \label{gamma}
\end{equation}
where $\sigma$ is the cross section for collisions between ThO and the buffer gas, $\bar{v}_{0,\mathrm{b}}$ is the mean thermal velocity of the buffer gas atoms (the subscript $0$ indicates in-cell, stagnation quantities), and $L_\mathrm{cell}$ is the length of the cell.  The  parameter $\gamma$ determines whether the beam system is running with hydrodynamic entrainment or not.  When $\gamma_\mathrm{cell}\gtrsim 1$ the molecules are entrained in the flow of the buffer gas, which results in an order of unity fraction of cooled, in-cell molecules being extracted into the molecular beam.  For a hydrodynamic source, during the initial cooling and through the extraction phase, the molecules are in a low-density environment colliding only with inert gas atoms.  These aspects of hydrodynamic buffer gas beams make this method an attractive alternative for species that are reactive, refractory, or otherwise difficult to obtain in the gas phase.

In the beam region, just outside the cell aperture, the molecules undergo collisions with buffer gas atoms. Because the average velocity of the buffer gas atoms is higher than that of the (typically) heavier molecules, and because the collisions are primarily in the forward direction, the molecules are accelerated to a forward velocity, $v_f$, which is larger than the thermal velocity of the molecules. As the stagnation density is increased, $v_f$ increases until it approaches a maximum value, which is that of the forward velocity of the buffer gas.  In flow regimes considered in the existing literature, the transverse spread is given approximately by a  Maxwell-Boltzmann distribution with temperature equal to the cell temperature, since there are typically not enough collisions outside the cell to increase this spread.  In this model the molecular beam is therefore a thermal cloud of molecules with a center of mass motion given approximately by the forward velocity of the buffer gas.  The angular distribution has a characteristic full (apex) angle $\theta$ given  by $\tan(\theta/2)=v_\perp/v_{||}\approx \bar{v}_{0,\mathrm{mol}}/\bar{v}_{0,\mathrm{b}}\approx\sqrt{m_\mathrm{b}/m_\mathrm{mol}}$, where the subscripts b and mol represent the buffer gas and molecule, respectively.  Since typically $m_\mathrm{mol}\gg m_\mathrm{b}$, we have $\theta\ll 1$ so $\theta\approx 2\sqrt{m_\mathrm{b}/m_{mol}}$.  The characteristic solid angle is then $\Omega=2\pi(1-\cos(\theta/2))\approx \pi\theta^2/4 = \pi m_\mathrm{b}/m_{mol}$.  This quantity is approximately 0.1 and 0.3 for ThO in He and Ne, respectively.  This model accurately describes the results of previous buffer gas beam experiments \cite{Maxwell2005,Patterson2007,Prd09}. As will be seen in section \ref{results}, we have studied beam operation at higher Reynolds numbers than previous experiments and therefore additional effects must be considered.

\subsubsection{Beam parameterization}

Buffer-gas beams typically operate in the intermediate regime between effusive and supersonic, where both molecular kinetics and fluid-like behavior are important. The parameters determining the operating regime of the beam dynamics outside the cell are the Reynolds and Knudsen numbers.  The Knudsen number is defined as $Kn=\lambda/d = 1/\sqrt{2}n_0\sigma d$, where $\lambda$ is the mean free path of the buffer gas and $d$ is a characteristic length scale, in our case the cell aperture diameter $d_a$.  The Knudsen number is related to the Mach number, $M\!a$, and Reynolds number, $Re$, by \cite{Sone2007}
\begin{equation}
\frac{1}{2}(Kn)(Re) \approx M\!a
\end{equation}
Near the cell aperture, the buffer-gas atoms are traveling at approximately their thermal velocity $\bar{v}_{0,\mathrm{b}}=\sqrt{8k_BT_0/\pi m_\mathrm{b}}$, where $k_B$ is Boltzmann's constant.  For the purposes of our qualitative estimates, this velocity is near to the speed of sound $c_{0,\mathrm{b}}=\sqrt{5k_BT_0/3m_\mathrm{b}}=0.8\bar{v}_{0,\mathrm{b}}$. Therefore $M\!a\approx 1$ near the cell aperture, and we have the relation
\begin{equation}
\frac{1}{2}(Kn)(Re) \approx 1. \label{knre}
\end{equation}

Using the formulas above, we estimate the relationship between flow and Reynolds number to be
\[ Re\approx0.7\times(f_0/1\textrm{ SCCM})\times(d_\mathrm{a}/4.5\textrm{ mm}), \]
where we estimate the collision cross section (from diffusion measurements) to be $\sigma \approx3\times10^{-15}$ cm$^{2}$.

Buffer gas beams typically operate in the regime $Kn \approx 1-10^{-2}$ \cite{Krems2009}.  Effusive beams require $Kn>1$ so that the aperture does not alter the properties of the atoms extracted from the cell, while supersonic beams typically have $Kn\lesssim 10^{-3}$ \cite{Scoles1988}.

\section{Measured Beam Properties}\label{results}

We studied the beam with a variety of buffer gas flows, aperture sizes, and cell temperatures for both helium and neon buffer gases using continuous wave laser spectroscopy from the ThO ground electronic state $X$ ($v=0$, $\Omega=0^+$, $B_e$=0.33 cm$^{-1}$) to the excited electronic state $C$ ($v=0$, $\Omega=1$, $T_0=14489.90$ cm$^{-1}$, $B_e=0.32$ cm$^{-1}$) \cite{Esa65} at 690 nm.   Two diode lasers are each locked to a stabilized frequency source via a Fabry-P\'{e}rot transfer cavity.  The frequency of one laser is scanned to obtain spectra, while the other is kept at a fixed frequency and used to normalize against variation in the ablation yield (typically a few percent from shot-to-shot).  The scanning laser detuning, Nd:YAG pulses, and data acquisition are all synchronized via a master control computer.  The scanning laser is split into multiple beams and used for absorption transverse to the molecular beam at several distances after the cell aperture, and for laser-induced fluorescence (LIF) parallel to the molecular beam as shown in figure \ref{fig:apparatus}.  Absorption data is obtained using silicon photodiodes, and laser-induced fluorescence is collected with either a CCD camera or a photomultiplier tube (PMT).

\subsection{Rotational Cooling}

The rotational temperature was determined by fitting a Boltzmann distribution to the lowest six rotational levels ($J=0$ to $J=5$) of the ground state $X$.  Population was determined from the optical density of absorption on the $X$ to $C$ transition.   The lines $R(0)$, $Q(1),\ldots,Q(5)$ were used to obtain the population in $X,J=0,1,\ldots,5$ respectively.  Rotational temperatures with both helium and neon buffer gases are shown in figure \ref{fig:rot_pop_fits}. The minimum measured rotational temperature, measured 6 cm from the cell, as a function of buffer gas flow and aperture size, was $2.0\pm 0.8$ K with neon buffer gas, and  $1.7\pm0.5$ K with helium buffer gas.  These represent an increase of a factor of 8.2 and 2.8 in the $X,J=0$ population with neon and helium buffer gas, respectively, from the distribution present at the cell temperature.

With neon buffer gas the rotational temperature decreases with both increasing flow and increasing distance from the cell aperture.  The rotational temperature does not change after a distance of 2 cm after the cell, indicating that the cooling collisions have stopped before this distance.  With helium buffer gas the rotational temperature is largely independent of flow, distance from the cell, and aperture size as measured with 14 different flows, three different distances after the cell, and three different aperture sizes.  The temperature of the molecules just outside of the cell, however, is lower than the cell temperature, even for the lowest flow and largest aperture.  This behavior is unexpected: at the lowest flow (1 SCCM) and largest aperture (4.5 mm side square), the flow regime is effusive $(Kn\sim 1)$ and so we should see no additional cooling below the cell temperature of $\sim 5$ K.  More low-flow helium phenomena, along with possible explanations, are discussed in section \ref{sec:lowflow}.

\begin{figure}[htbp]
	\centering
		\includegraphics[width=0.8\textwidth]{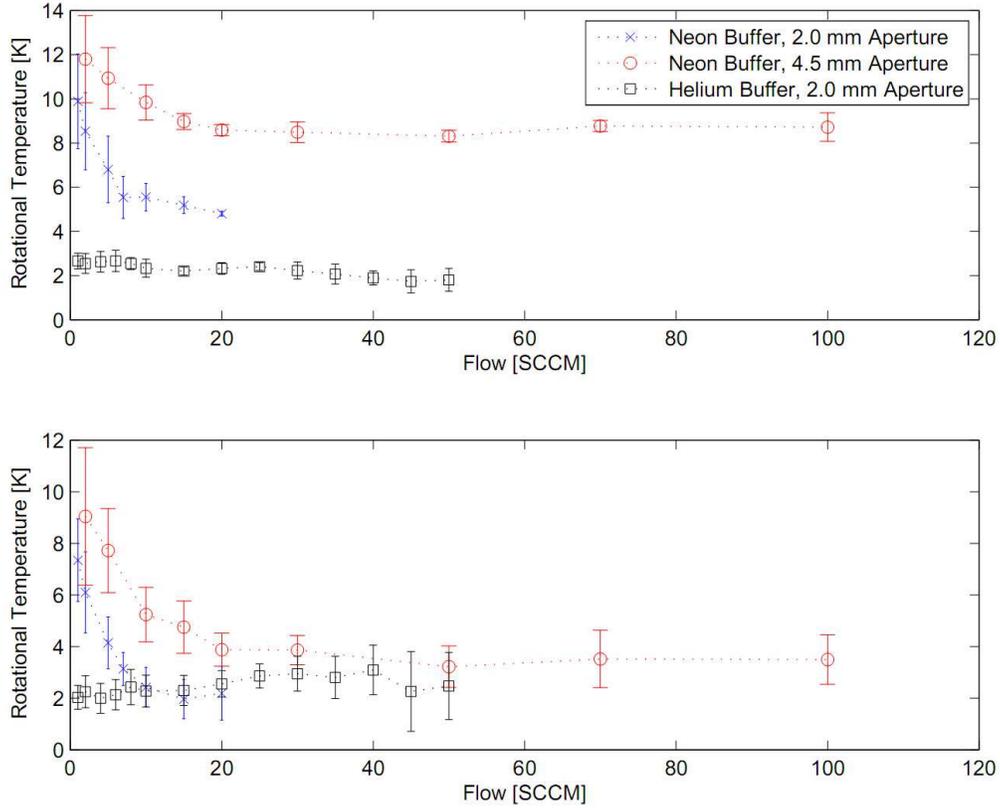}
	\caption{Rotational temperatures in the buffer gas beam.  Top: temperatures measured at the cell aperture.  Bottom: final rotational temperatures in the expansion, measured 6 cm after the cell aperture.  The cell temperatures are 5$\pm 1$ K and 18$\pm 1$ K for helium and neon buffer gases, respectively.  Error bars represent the 95\% confidence interval of the Boltzmann fit of a single data run.  Other data runs fell within error bars reported here.}
	\label{fig:rot_temps}
\end{figure}

\begin{figure}[htbp]
	\centering
		\includegraphics[width=0.6\textwidth]{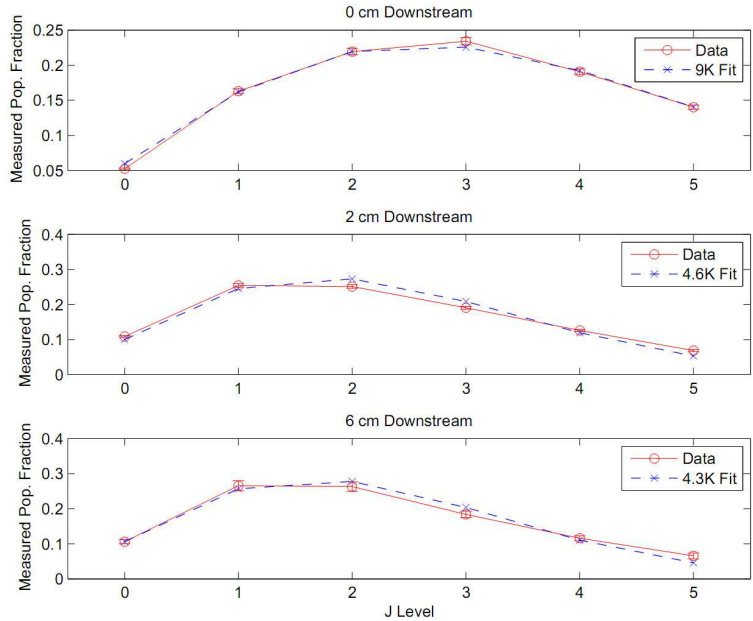}
	\caption{Rotational level populations with Boltzmann distribution fits for 30 SCCM neon flow, 4.5 mm aperture.  The top, middle, and bottom plots show the distributions near, 2 cm from, and 6 cm from the cell aperture, respectively.}
	\label{fig:rot_pop_fits}
\end{figure}

For both buffer gases, the molecules approach some minimum rotational temperature that does not decrease with additional flow.  This minimum temperature appears to be about 2-3 K, and is similar for both helium and neon.  In fact, the rotational temperature just outside a 9 K cell with a helium buffer gas flow has a similar minimum temperature of $2.9\pm1.3$ K.

\subsection{Forward velocity}\label{sec:vf}

The forward velocity of the ThO beam was measured at distances between about 6 cm and 16 cm from the cell aperture using laser-induced fluorescence imaging.  A counter-propagating, red-detuned pump beam excites the molecules on the $Q(1)$ or $Q(2)$ line of the $X-C$ transition, and the fluorescence is collected with a CCD or PMT.  The camera gives spatial information about the beam, but since the exposure time is longer than the molecular pulse duration the camera averages over an entire pulse.  To get time-dependent information, we use the PMT.  The forward velocity is determined by fitting a gaussian to the longitudinal fluorescence spectrum and comparing the center to that of a transverse absorption spectrum. 

Values for the forward velocity are between 120 and 200 m/s, which are slower than typical supersonic expansions.  The final velocity of a supersonic expansion of a monoatomic carrier gas is given by \cite{Pauly2000}
\begin{equation}
v_{||}^* = \sqrt{\frac{5}{2}\frac{2k_BT_0}{m}} \approx 1.6 v_{p,0} \label{finalvf},
\end{equation}
where $m$ is the carrier gas particle mass, $T_0$ is the stagnation temperature in the source, and $v_{p,0}$ is the most probable thermal velocity of the carrier in the source.  From this equation, the final velocity is about 600 m/s for a supersonic expansion of room-temperature argon, or about 300 m/s for an expansion of 210 K xenon.  With neon buffer gas in an 18 K cell, the forward velocity approaches a value that is very close to the final velocity of $\approx 200$ m/s predicted by equation (\ref{finalvf}).  With helium buffer gas, the mean forward velocity approaches about 70\% of the value $230$ m/s predicted by equation (\ref{finalvf}) for a 5 K cell.  The lower-than-expected value for the helium-cooled beam is likely due to collisions with the background helium that accumulates due to the limited pumping speed of the activated charcoal, which is not a problem with neon because once it sticks to a 4 K surface, it remains there nearly indefinitely (i.e. has negligible vapor pressure at 4 K).  Increasing the amount of charcoal and improving its placement mitigated this problem; however, the issue remained present.  Further discussion is presented in section \ref{sec:hevsne}.  Interpretation of the helium forward velocity data is additionally complicated by the fact that, unlike with the case of neon buffer gas, it varies in time over the molecule pulse duration, as shown in figure \ref{fig:he_forward_velocity_vs_time}.  This dependence of beam properties on time after ablation is perhaps due to the finite amount of time required to thermalize the ThO molecules in the buffer gas cell, which is much smaller with neon due to neon's smaller mass mismatch with ThO, and the fact that the heat introduced by ablation results in a smaller fractional change in temperature at 18 K versus 5 K.  Additionally, the forward velocity with helium buffer gas varies by as much as $\sim10$\% if the ablation spot is moved, and as the charcoal cryopumps become full of helium and the pumping speed changes, as discussed in section \ref{sec:hevsne}.  These effects are not observed with neon buffer gas.

\begin{figure}[htbp]
	\centering
		\includegraphics[width=0.8\textwidth]{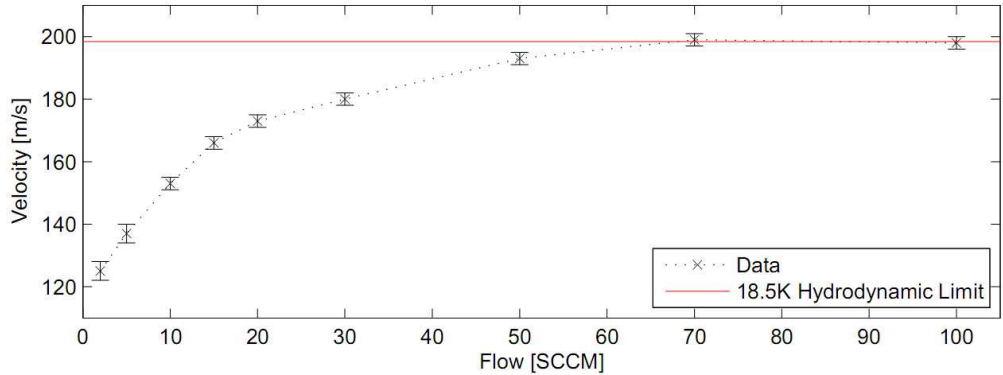}
	\caption{Mean forward velocity vs. flow with neon buffer gas.  The solid line is the hydrodynamic limit for the forward velocity of neon atoms exiting an 18.5 K cell, given by equation (\ref{finalvf}).  The forward velocity of the molecules with neon buffer gas cooling varies by no more than 10\% over a pulse.  Error bars represent the 95\% confidence interval of spectral fits.}
	\label{fig:forward_velocity}
\end{figure}

\begin{figure}[htbp]
	\centering
		\includegraphics[width=0.8\textwidth]{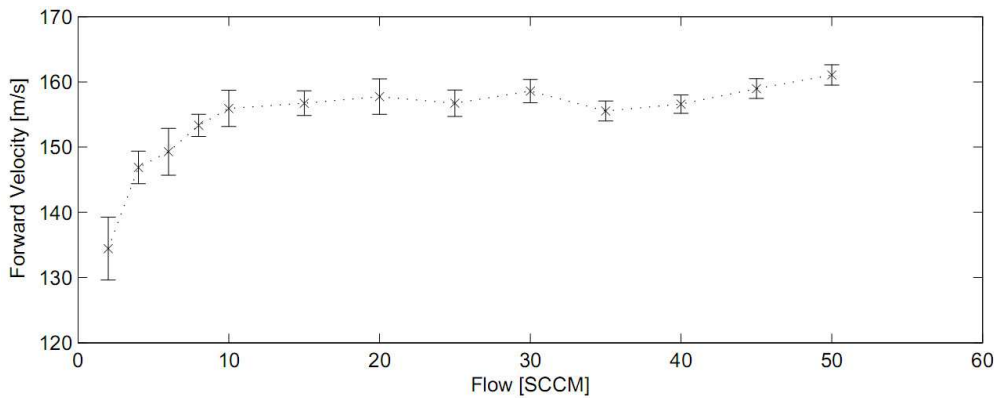}
	\caption{Mean forward velocity vs. flow with helium buffer gas.  Error bars represent the 95\% confidence interval of spectral fits.}
	\label{fig:he_forward_velocity}
\end{figure}

\begin{figure}[htbp]
	\centering
		\includegraphics[width=0.8\textwidth]{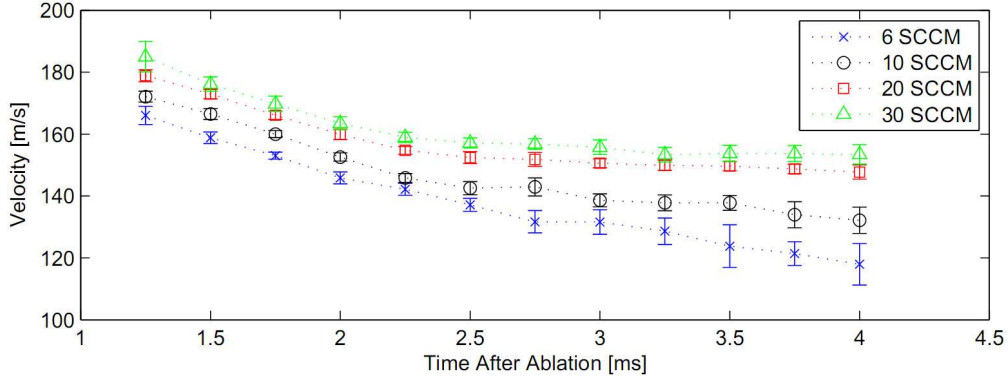}
	\caption{Forward velocity vs. time for several different flows of helium.  Error bars represent the 95\% confidence interval of spectral fits.}
	\label{fig:he_forward_velocity_vs_time}
\end{figure}

\subsubsection{The velocity vs. flow relationship}

We can model the shape of the velocity vs. flow curve with the ``sudden freeze" model \cite{Pauly2000}, in which we assume that the molecules are in equilibrium with the buffer gas until some distance where the density is low enough that the gases decouple, and the molecular beam properties are frozen.  The velocity of the buffer gas scales with the density $n$ as \cite{Pauly2000}
\[ v(x) \approx 1.6 v_{\mathrm{p},0}\left[1-\left(\frac{n(x)}{n_0}\right)^{2/3}\right]^{1/2}, \]
where $v_{p,0}=(2k_BT_0/m_\mathrm{b})^{1/2}$ is the most probable velocity of the buffer gas in the cell ($\approx 120$ m/s for 17 K Ne), $k_B$ is Boltzmann's constant, $x$ is the distance from the cell aperture, $n_0$ is the cell stagnation density, and the collision cross section is estimated to be $\sigma \approx3\times10^{-15}$ cm$^{2}$.  Introducing the normalized distance $\xi\equiv x/d_\mathrm{a}$, the far-field number density scales as $n(\xi)/n_0\approx 0.2\xi^{-2}$ \cite{Pauly2000}, so $v(\xi)\approx 1.6 v_{\mathrm{p},0}\sqrt{1-0.3\xi^{-4/3}}$.  For a monoatomic hard-sphere gas, the location where collisions freeze is given by \cite{Pauly2000}
\begin{equation}
\xi_0 \approx \left(0.1 \sigma n_0 d_\mathrm{a}\right)^{3/5} \approx \left(0.4 \frac{\sigma f_0}{d_\mathrm{a} v_{\mathrm{p},0}}\right)^{3/5} \approx 0.2\left(\frac{f_0}{1\textrm{ SCCM}}\right)^{3/5} \label{xi0}
\end{equation}
 where in the last equality we used $d_\mathrm{a}=4.5$ mm.  The 3/5 exponent is model dependent and should be considered approximate.  Therefore, if we assume that the molecules and buffer gas are in equilibrium until the position $\xi_0$, the final molecule velocity will be given by
\[
v(\xi_0) \approx 1.6 v_{\mathrm{p},0}\left[ 1- 3  \left(\frac{f_0}{1\textrm{ SCCM}}\right)^{-4/5} \right]^{1/2}
\]

\begin{figure}[htbp]
	\centering
		\includegraphics[width=0.80\textwidth]{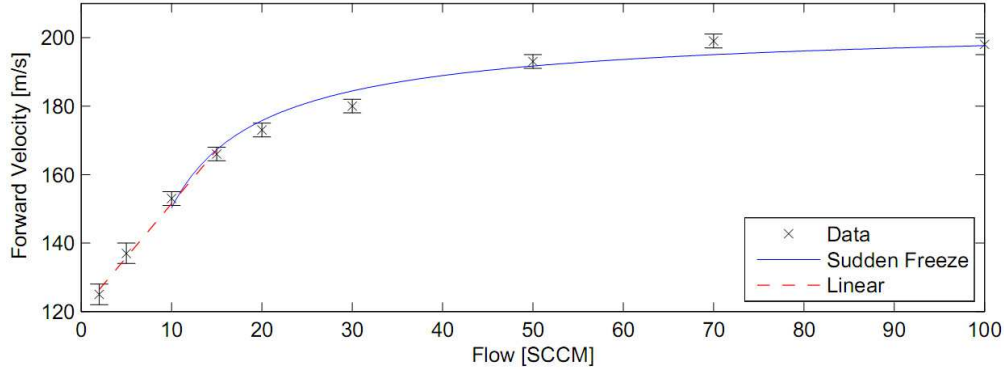}
	\caption{Forward velocity for neon buffer gas with a 4.5 mm aperture, along with a linear fit at low flow and a sudden freeze fit at high flow.  The sudden freeze fit is to the function $v = \alpha\sqrt{1-\beta f^{-4/5}}$, with parameters $\alpha=205\pm 6$ m/s, $\beta = 3.0\pm 0.5$.  The parameters are close to the expected values of $\alpha\approx 200$ m/s (the final velocity) from experiment, and $\beta\approx 3$ from the estimate above.  The linear model has a slope of 3.1$\pm$0.8 (m/s)/SCCM, in good agreement with our estimate of 4.5 (m/s)/SCCM.}
	\label{fig:ne_forward_velocity_piecewise_fit}
\end{figure}

This sudden-freeze model is valid when there are collisions in the beam, \emph{i.e.} when $\xi_0\gtrsim 1$, which from equation (\ref{xi0}) occurs at approximately 15 SCCM neon flow with the 4.5 mm aperture.  Below this flow we are in the regime where the forward velocity increases linearly \cite{Maxwell2005}, so we can fit a line to the flows $\leq 15$ SCCM and the sudden freeze model for flows $\geq$ 15 SCCM, as shown in figure \ref{fig:ne_forward_velocity_piecewise_fit}.  

To estimate the slope of the linear part, we use the model from \cite{Maxwell2005}; near the aperture, the molecules undergo approximately $d_\mathrm{a}/\lambda_0\approx Re/2$ collisions (from equation (\ref{knre})).  Each of these collisions gives the molecules a momentum kick in the forward direction of about $\delta p_\mathrm{mol}\approx m_\mathrm{b} v_\mathrm{b}$, so the net velocity boost is $\Delta v_{mol} \approx (d_\mathrm{a}/\lambda_0)\delta p_\mathrm{mol} \approx (Re) v_\mathrm{b} m_\mathrm{b}/2m_{mol}$.  The forward velocity of the buffer atoms is $1.3 v_{p,0}\approx 160$ m/s, which is the mean forward velocity of an effusive beam \cite{Scoles1988}).  The estimated conversion from flow to Reynolds number is given by $Re\approx0.7\times(f_0/1\textrm{ SCCM})\times(d_\mathrm{a}/4.5\textrm{ mm})$, so the velocity increase is estimated to be approximately given by $\Delta v_{mol}\approx 4.5\times f_0/$(1 SCCM).  A fit with this model is shown in figure \ref{fig:ne_forward_velocity_piecewise_fit}, where we find good agreement with our estimated parameters.

\subsection{Divergence}

The transverse absorption spectra give information about the angular spread of the molecular beam.   We divide the full width at half maximum (FWHM) of the transverse velocity distribution by the forward velocity to obtain $\theta_{FWHM}$, the FWHM of the angular distribution.  The angle $\theta_\text{FWHM}$ is related to the half-maximum solid angle $\Omega$ by $\Omega=2\pi(1-\cos(\theta_\text{FWHM}/2))$.

For neon buffer gas, the divergence has a minimum at $\Omega\approx 0.35\pm0.03$ $(\theta_\textrm{FWHM}\approx 39^\circ)$, close to the hydrodynamic entrainment prediction of $\pi m_\textrm{Ne}/m_\textrm{ThO}\approx 0.3$.  The location of the minimum divergence appears at approximately the same Reynolds number for different apertures; however, as discussed in section \ref{sec:highflow}, the neon signals at smaller aperture are greatly reduced, and therefore the minimum cannot be determined as accurately.

For helium buffer gas, the minimum divergence of about $\Omega\approx 0.22\pm 0.06$ $(\theta_\textrm{FWHM}\approx 30^\circ)$ occurs for 2 SCCM flow, and then steadily increases to $0.29\pm0.07$ at 50 SCCM.  The smallest measured divergence was $\Omega\approx 0.17\pm 0.06$ $(\theta_\textrm{FWHM}\approx 27^\circ)$ with a 3.0 mm aperture; however, changing the aperture size does not have an appreciable effect on the divergence vs. flow relationship.  

\begin{figure}[htbp]
	\centering
		\includegraphics[width=0.8\textwidth]{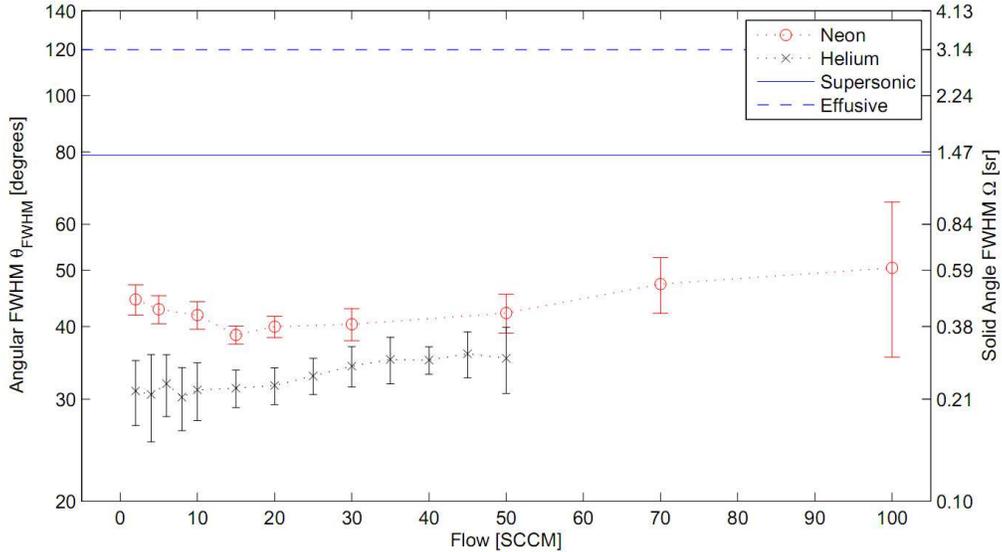}
	\caption{Divergence in the beam with a 4.5 mm aperture, measured 2 cm after the cell aperture.  The cell temperatures are $18\pm 1$ K for neon, $5\pm 1$ K for helium.  For this aperture size, the conversion from flow to Reynolds number is $Re\approx 0.7\times($flow/(1 SCCM)).  The divergences for a supersonic and effusive beam come from \cite{Scoles1988}.  Error bars represent the 95\% confidence interval from spectral fits.}
	\label{fig:divergence}
\end{figure}

The spectra analyzed in this section were obtained by integrating the optical density (OD) vs. time after the YAG pulse and varying the absorption laser detuning.  Integrating over the entire pulse yields the mean transverse spread of all the molecules.  Alternatively, we can look at the transverse width as a function of time after the ablation pulse.  With neon buffer gas the transverse spread does not change by more than 10\% over the duration of a single pulse of molecules, but with helium the transverse spread changes by as much as 30\%.  The divergence shown in figure \ref{fig:divergence} was calculated using the mean spectral width, defined as the spectral width of the entire molecule pulse.  As with the forward velocity data, the transverse velocity spread with helium buffer gas varies by as much as $\sim10$\% if the ablation spot is moved, and as the charcoal cryopumps become full of helium; however, the data typically falls within the error bars shown in figure \ref{fig:divergence}. These effects are not observed with neon buffer gas.

\subsubsection{The divergence vs. flow relationship}

Here we present a model to describe how the beam divergence varies with flow. This model is similar to those discussed in \cite{Anderson1967,FernandezdelaMora1988}.  For low flows we are in the regime where forward-peaked collisions with the buffer gas near the aperture result in a forward velocity proportional to flow, but the transverse spread is fixed, as described in section \ref{buffertheory}.  The result is a linearly decreasing divergence with increasing flow.  For high flows, the molecules are fully boosted to the forward velocity of the buffer gas.  Additionally, there are an increasing number of collisions in the beam region between the ThO molecules and the buffer gas.  Since the thermal distribution of the buffer gas defines its transverse velocity spread, which is larger than that of the molecules by a factor of $\sqrt{m_{mol}/m_\mathrm{b}}$, collisions with the buffer gas in the beam increase the transverse velocity spread and therefore increase the divergence of the molecules.  In this regime, the molecules outside of the aperture no longer constitute a thermal cloud as they do in the low flow case.  Absorption spectra from lasers perpendicular to the molecular beam no longer give temperature information, but instead give the angular spread of the beam profile. 

With a simple model, we can estimate the slope of the angular spread vs. flow curve.  Near the aperture, the buffer gas atoms must follow a convergent trajectory since the flow cross-section narrows from the cell diameter $d_\mathrm{cell}\approx 13$ mm to the aperture diameter $d_\mathrm{a}\approx 5$ mm.  The typical mean buffer gas velocity normal to the beam axis is given approximately by the flow velocity.  The relationship between the in-cell flow velocity $v_\mathrm{cell}$ and the flow velocity in the aperture, which is approximated by $v_{p,0},$ is given by $d_\mathrm{cell}^2 v_\mathrm{cell}\approx d_\mathrm{a}^2 v_{p,0}$, or $v_\mathrm{cell}\approx v_{p,0}(d_\mathrm{cell}/d_\mathrm{a})^2\approx 18$ m/s.  There are $\approx Re/2$ collisions near the aperture, so the change in the transverse spread is given by $\Delta v_{mol,\perp} \approx (Re/2) v_\mathrm{cell} m_\mathrm{b}/m_{mol}\approx (0.7 \textrm{ m/s})\times Re$, in good agreement with the typical data fit value of (0.4 m/s) $\times$ $Re$, measured for several aperture sizes.  The transverse spread downstream displays similar linear increases, however modeling is complicated by the expansion dynamics.  Typical slopes for the transverse spread vs. Reynolds number relationship are $\sim 1$ (m/s)/($Re$).

\begin{figure}[htbp]
	\centering
		\includegraphics[width=0.8\textwidth]{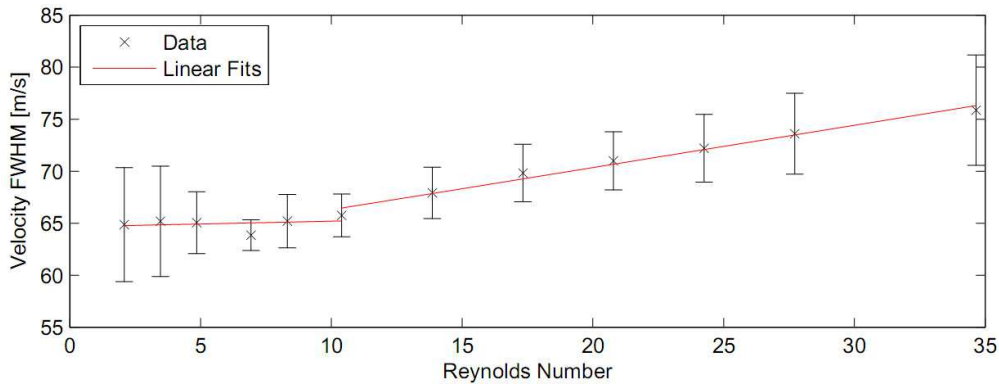}
	\caption{The transverse spectral width vs. Reynolds number with neon buffer gas and a 4.5 mm cell aperture, measured at the aperture.  At low flows the transverse widths inside the cell and outside the cell are the same, and indicate a thermal distribution of $T\approx 20$ K.  The slope of the fit line at these low flows is $(0.06\pm0.27)$ m/s, consistent with zero.  At higher flows the transverse width in the beam begins to linearly increase, with slope $(0.41\pm0.06)$ m/s.  Note that the transition Reynolds number agrees with that from figure \ref{fig:ne_forward_velocity_piecewise_fit}.  Error bars represent the 95\% confidence interval of spectral fits.} 
	\label{fig:trans_width}
\end{figure}

\subsection{Translational Cooling}\label{transcool}

\begin{figure}[htbp]
	\centering
		\includegraphics[width=0.8\textwidth]{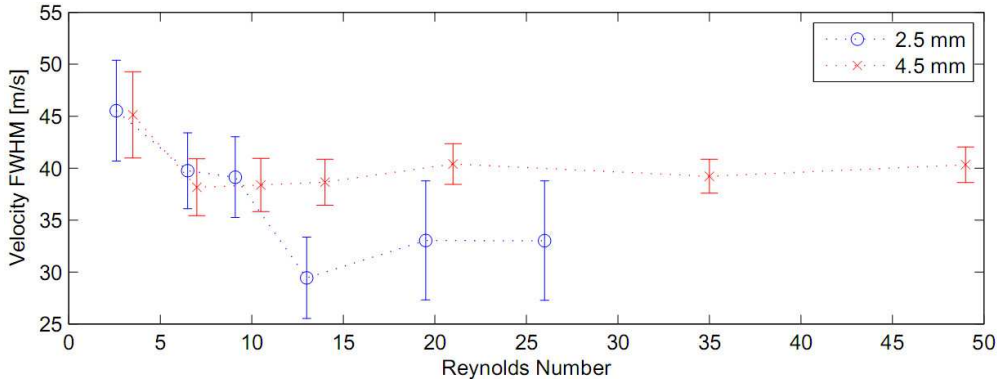}
	\caption{The velocity spread in the forward direction with neon buffer gas.  Error bars represent the 95\% confidence interval of spectral fits.}
	\label{fig:ne_forward_spread}
\end{figure}

The translational temperature in the longitudinal direction was measured by collecting laser-induced fluorescence after excitation from a laser beam counter-propagating with respect to the molecular beam, as described in section \ref{sec:vf}.  With neon buffer gas, the lowest translational temperature (in the center of mass frame of the molecular pulse) was $4.4\pm1.2$ K, corresponding to a velocity spread of $29\pm 4$ m/s.  Increasing the aperture diameter increases the final velocity spread, as shown in figure \ref{fig:ne_forward_spread}.

\begin{figure}[htbp]
	\centering
		\includegraphics[width=0.8\textwidth]{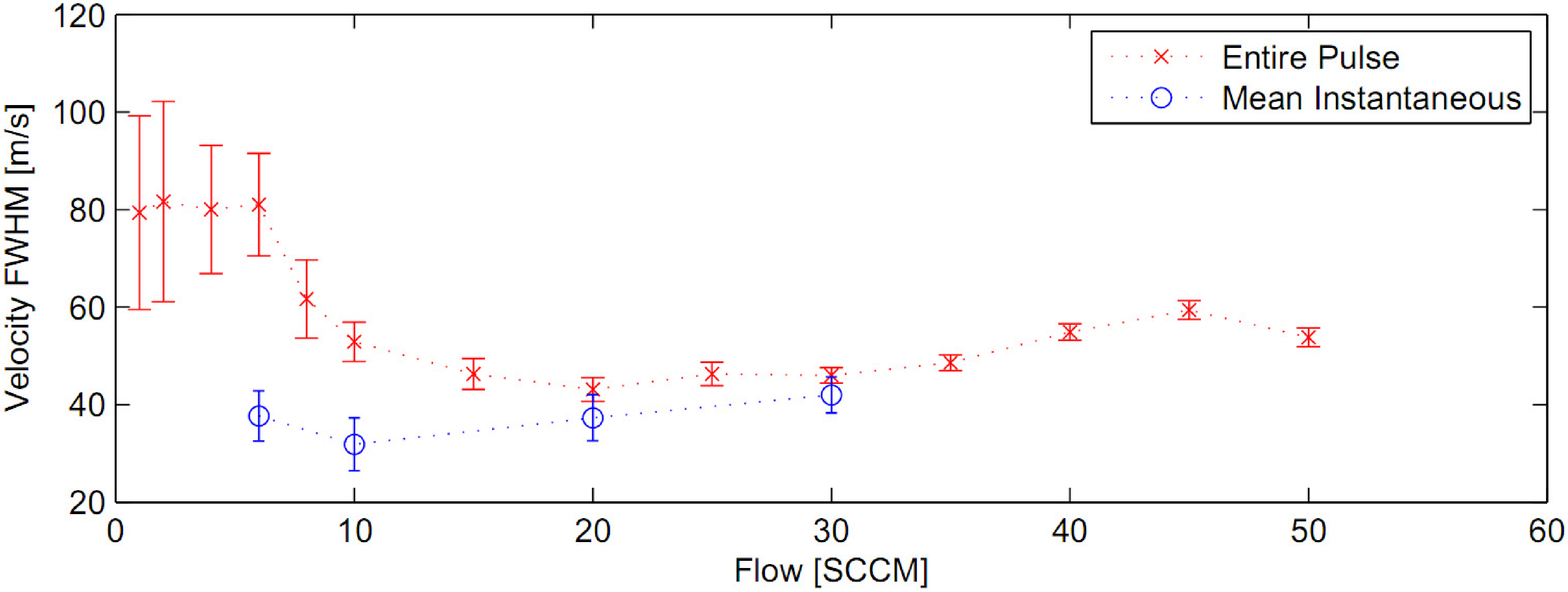}
	\caption{The velocity spread in the forward direction with helium buffer gas.  Error bars represent the 95\% confidence interval of spectral fits.  The data points marked by the $\times$ symbols indicate velocity spread averaged over the entire molecule pulse, while the data points marked by the $\circ$ symbols indicate the mean instantaneous velocity spread obtained from time-resolved fluorescence data.}
	\label{fig:he_forward_spread}
\end{figure}

With helium buffer gas the data interpretation is more difficult because the spectral properties of the beam change as a function of time after ablation.  Similar to the analysis for the helium forward velocity in section \ref{sec:vf}, we can measure the time-resolved fluorescence from the molecules to extract spectral properties as a function of time after ablation.  Figure \ref{fig:he_forward_spread} shows the forward velocity spread for the entire molecule pulse, and the mean instantaneous forward velocity spread extracted from the time-resolved spectra.  The instantaneous velocity spread is comparable to that with neon; however, the changing forward velocity of the beam (see figure \ref{fig:he_forward_velocity_vs_time}) results in the forward velocity spread of the entire pulse being much larger.  With neon, neither the forward velocity nor the forward velocity width change by more than 10\% over the duration of the pulse.

\subsection{Cell Extraction and Molecule Production}

\begin{figure}[htbp]
	\centering
		\includegraphics[width=0.8\textwidth]{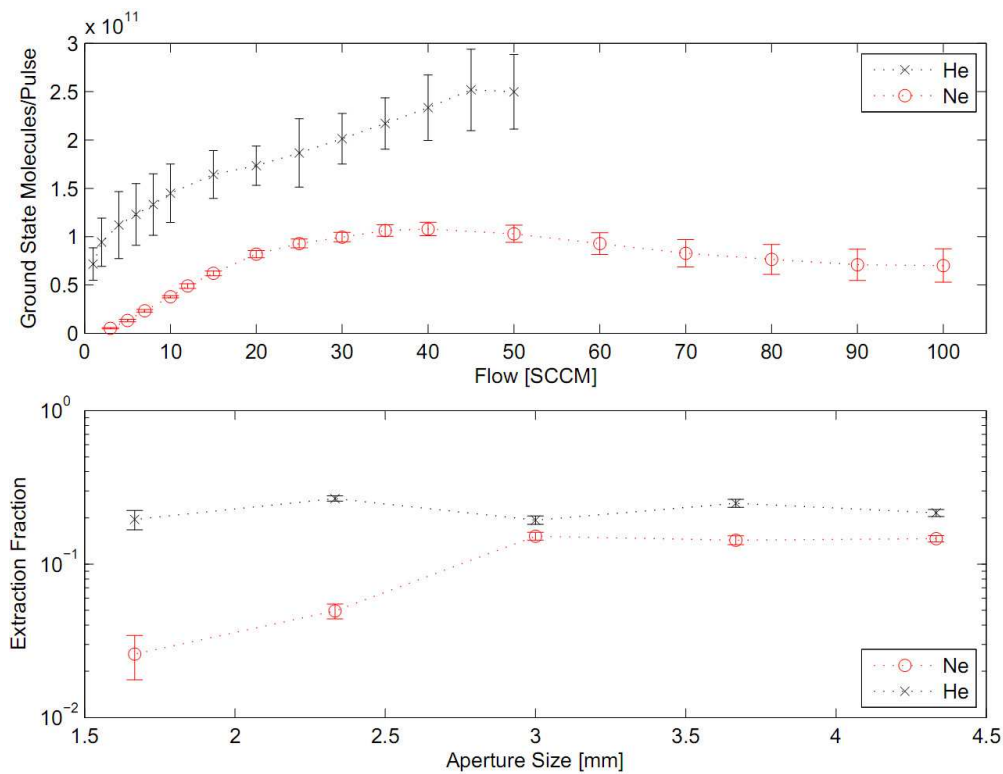}
	\caption{Top: Number of absolute ground state molecules output per pulse with helium and neon buffer gases, using a 4.5 mm cell aperture.  Bottom: Fraction of molecules extracted out of the cell into the beam, with neon and helium buffer gases.  Error bars represent the 95\% confidence interval from spectral fits.}
	\label{fig:total_molecules_and_extraction}
\end{figure}

We varied the flow and aperture size with a fixed cell geometry and measured the molecule output from the cell.  When the flow is large enough to ensure good extraction from the cell (\emph{i.e.} $\gamma_\mathrm{cell}\gtrsim 1$), typical molecule outputs are $\sim 10^{11}$ in the $X,J=0$ absolute ground state, or $\sim 10^{12}$ total molecules (in all states) per pulse, estimated from absorption spectra taken immediately after the cell aperture.  The optical absorption cross section is estimated from the $C$ state radiative decay lifetime and the calculated \cite{Wentink1972} Franck-Condon factors for the transition.  A lower bound of $480$ ns for the $C$ state lifetime was measured by exciting molecules into the $C$ state with resonant pump light, and then measuring the fluorescence as the molecules decay back into $X$ after rapidly extinguishing the pump light \cite{spectroscopypaper}.  Since the temporal width of the pulse is $\sim 1$ ms, the peak instantaneous output rate is $\sim 10^{14}$ molecules per second per state in $J=0$, or $\sim 10^{15}$ total number of molecules per second.  Behavior of the number of molecules output per pulse is plotted in figure \ref{fig:total_molecules_and_extraction}.  With neon buffer gas, the flow-output behavior is very similar for different aperture sizes larger than about 3.0 mm, and has a maximum output around $\gamma_\mathrm{cell}\approx 1-2$.  With helium buffer gas, the shape of the curve is non-repeatable; specifically, the shape changes if the YAG ablation spot is moved, which is not the case with neon.  However, the approximate number output is typically within a factor of two of the data in figure \ref{fig:total_molecules_and_extraction} for all conditions.

According to the simple hydrodynamic entrainment theory, the extraction of molecules from the cell is governed by the parameter $\gamma_\mathrm{cell}$ from equation (\ref{gamma}).  This parameter does not have an explicit dependence on the aperture size; however, we find that there can be a strong dependence on aperture size that varies with gas type.  For neon buffer gas, the extraction fraction is constant for cell apertures larger than about 3 mm, then falls off rapidly with decreasing aperture size.  For helium buffer gas, the maximum extraction fraction does not depend significantly on the aperture size.  It should be noted that based on previous observations, both published \cite{Maxwell2005,Patterson2007} and unpublished, the cell extraction is typically dependent, sometimes in puzzling ways, on species, ablation properties, internal cell geometry, and collimation geometry.

\subsubsection{High-flow neon phenomena}\label{sec:highflow}

For very high neon flows (Reynolds number $\gtrsim$ 50) the beam behavior degrades.  The molecule pulse resulting from a single ablation shot begins to decompose into spatially and temporally variegated pulses with different spectral characteristics.  This is likely a result of the post-ablation dynamics in the high density buffer gas inside the cell, as we have observed the onset of similar behavior occurring in high-density, ablation-loaded buffer gas cells in other species \cite{otherspecies} in the past.  Fortunately, this behavior occurs for flows significantly above the point where the cell extraction is maximum, and therefore is not near the optimal flow for most conceivable beam-based spectroscopy experiments. 

\subsubsection{Low-flow helium phenomena}\label{sec:lowflow}

For low flows of helium the behavior of the beam is unusual and not in good agreement with other helium buffer gas beam experience \cite{Maxwell2005}.  At low flow there should be little extraction and poor thermalization; however, in the data presented here we have evidence of a high-extraction, well-thermalized beam at flows as low as 1 SCCM with aperture size as large as 4.5 mm, or $Re\approx$ 1 and $\gamma_\textrm{cell}<0.1$.  At this flow the mean free path in the cell is a few mm, so it is surprising that there is any significant thermalization or extraction.  A possible explanation for this behavior is that a helium film builds up in the cell (either on the walls or in the target) as helium gas flows into the cell between ablation pulses. Then, when the ablation pulse hits and the helium desorbs due to ablation heating, there is a pulse of higher buffer gas pressure at the moment the ThO is introduced into the cell.  Additionally, we observe that the phenomena disappear as the cell is heated, in which case we observe the expected \cite{Maxwell2005} increasing extraction with increasing flow.

\subsection{Comparison of Helium and Neon Buffer Gas Sources}\label{sec:hevsne}

Our data shows that a beam using neon buffer gas performs nearly as well as one using helium, but with much simpler technical requirements and a much larger cryopumping speed.  The minimum rotational temperatures with both helium and neon buffer gas differ by $<$1 K, despite the fact that the buffer gas cell sits at $\sim$5 K and $\sim$18 K for helium and neon respectively.  The forward velocity and divergence of the helium-based beam are slightly lower, however both can vary as much as 30\% over a single beam pulse, which is behavior not seen in our neon-based source.

In addition to time variations of beam properties within a single pulse, we find that the properties of helium-based beams also depend significantly on the location of the ablation spot, and on the pumping conditions of the buffer gas out of the beam region.  Helium in the beam region can only be pumped by a large surface area adsorbent, such as activated charcoal.  As the helium adsorbent fills up, the cryopumping speed begins to change, resulting in a significant change in beam properties.  To keep the beam properties consistent, the adsorbent must be emptied periodically (typically every few hours for our apparatus) by heating it and then pumping the helium out of the vacuum chamber.  In addition, we find that the performance of our helium-based beam is very sensitive to the amount and placement of both beam collimators and adsorbent: incorrect placement often results in the extinction of the molecular beam, as we have observed in our apparatus and several other similar apparatus.  Correct placement requires much trial and error, and is not currently understood, though has been achieved on other ThO beam test apparatus in our group.

Neon, on the other hand, is readily adsorbed by any 4 K surface, including neon ice, allowing for a cryopump of $>$1000 l/s in the beam region of our apparatus.  We have operated the neon-based molecular beam continuously, with 30 SCCM of neon flow, for over 24 hours with little increase in background pressure and no appreciable variation in beam properties.  The neon-based beam is also robust to variation of the collimator geometry.  We have experimented with several cell-collimator configurations, including changing the cell-collimator distance (\emph{in situ} via a motion feedthrough on the apparatus), and found that for all configurations with the neon-based beam, the collimator performed as expected and we have never had difficulty obtaining consistent, robust beam signal.  

Another advantage of a neon-based beam is that because the cell is kept at a higher temperature, the refrigerator cooling the cell can sustain a much higher heat load.  We have demonstrated operation of the neon-cooled ThO beam with nearly 10 W of input power from a 200 Hz pulsed YAG \cite{thelitron} and achieved stable production with single-shot yields comparable to those measured with a slower repetition rate, resulting in about a factor of 10 increase in time-averaged yield compared to the data presented earlier in this paper.  Further characterization of the beam with the high repetition rate is ongoing.

\subsection{Comparison to Other Beam Sources}\label{sec:otherbeams}

Table \ref{tab:comparison} shows a comparison between the two buffer gas beam sources discussed in this paper and two common beam types, supersonic and effusive.  A supersonic beam of ThO has been created in the past \cite{Goncharov_etal2005}, however a thorough characterization of beam properties was not performed so we instead compare to a beam of a similar (refractory and chemically reactive) species, YbF \cite{Ths+02}.  The effusive beam properties are estimated from ThO vapor pressure data \cite{Ackermann1973}.  Which beam properties are most important depends on the application, but for a proposed \cite{Vutha2010} precision spectroscopy experiment, the high count rate afforded by the increased brightness, and the long interaction time afforded by the low forward velocity, makes a buffer gas source an ideal choice.

\begin{table}[htbp]
	\centering
		\begin{tabular}{l|lllllll}
			Beam Type  & Peak Brightness & Velocity & $T_{\mathrm{rotational}}$ & $T_{\mathrm{translational}}$ \\ \hline
			 Ne Buffer Gas (ThO, this work) & $3\times 10^{11}$/sr/pulse & 120-200 m/s & 2-9 K & 4-11 K \\
			He Buffer Gas (ThO, this work)  & $10\times 10^{11}$/sr/pulse & 120-160 m/s & 2-3 K &  5-10 K$^*$ \\
			Supersonic (YbF, \cite{Ths+02}) & $1\times 10^9$/sr/pulse & 290 m/s & 1 K & 1 K \\
			Effusive (ThO, predicted \cite{Ackermann1973}) &  $1\times 10^{11}$/sr/s & 540 m/s & 2000 K & 2000 K
		\end{tabular}
	\caption{Comparison of the buffer gas beam sources presented in this paper to a supersonic source and an effusive source.  Peak brightness is defined as the peak number of absolute ground state molecules per steradian per pulse (or per second for the continuous effusive source) in the beam.  The buffer gas sources are those presented in this paper, and the range of values given for the beam properties is the range of values measured with varying gas flow rates and aperture sizes.  The supersonic beam is a pulsed beam of YbF seeded in Xe from a 193 K source \cite{Ths+02}, and the effusive beam is an estimate for a 2000 K source based on ThO vapor pressure data \cite{Ackermann1973}.  $^*$Typical mean instantaneous spread; see \ref{transcool} for a discussion.}
	\label{tab:comparison}
\end{table}

\section{Conclusion}

We have demonstrated a general cryogenic buffer gas beam source operating in the intermediate flow regime with additional cooling from a free gas expansion, and studied its behavior in producing cold, slow, and bright molecular beams of ThO.  The flux, divergence, temperature, and forward velocity of the source compare very favorably to other beam technologies, especially for chemically reactive species.

\section{Acknowledgments}

Thanks to Stan Cotreau and Dave Patterson for technical assistance; and John Barry and Edward Shuman for helpful discussions.  This work was funded by a NIST Precision Measurement Grant and a grant from the National Science Foundation.

\bibliographystyle{ieeetr}
\bibliography{acme_beam_paper_2011}

\begin{thebibliography}{10}
\bibitem{charcoal}
{We used 8-30 mesh activated coconut charcoal from Spectrum Chemical, bonded to
  1/8" copper plates by Stycast 2850FT epoxy (with catalyst 24LV)}.
\bibitem{spectroscopypaper}
{A manuscript with a detailed description of the measurement is in
  preparation.}
\end{thebibliography}

\end{document}